\documentstyle[psbox]{WORKSHOP}

\begin{document}

\title{X-ray bursts and superbursts - recent developments}

% AUTHOR(S) 
\author{
Jean in 't Zand  % TO BE SPACED WITH ONE LINE
\\[12pt] 
 SRON Netherlands Institute for Space Research  \\
{\it E-mail: jeanz@sron.nl}
}

\abst{ The past decade and a half has seen many interesting new
  developments in X-ray burst research, both observationally and
  theoretically. New phenomena were discovered, such as burst
  oscillations and superbursts, and new regimes of thermonuclear
  burning identified. An important driver of the research with present
  and future instrumentation in the coming years is the pursuit of
  fundamental neutron star parameters. However, several other more
  direct questions are also in dire need of an answer. For instance,
  how are superbursts ignited and why do burst oscillations exist in
  burst tails? We briefly review recent developments and discuss the
  role that MAXI can play. Thanks to MAXI's large visibility window
  and large duty cycle, it is particularly well suited to investigate
  the recurrence of rare long duration bursts such as superbursts. An
  exploratory study of MAXI data is briefly presented.}

\kword{X-rays: bursts --- X-rays: binaries --- Stars: neutron}

\maketitle

\thispagestyle{empty}

\section{Introduction}

The first X-ray burst was detected in 1969 (Belian et al. 1972; see
also Kuulkers et al. 2009a), with the same instrumentation that
revealed gamma-ray bursts (Klebesadel et al. 1973) - the Vela 5 and 6
satellite series. Curiously, that first X-ray burst is still the
brightest ever recorded, with a peak flux close to 50 times the Crab
source. It is the brightest simply because it came from the
nearest-known X-ray burster: Cen X-4 at 1 kpc.

The real kick-off for X-ray burst research was in 1975, when the first
long-duration observations were performed with the first 3-axis
stabilized Dutch astronomy satellite ANS. Hunting for a black hole in
the globular cluster NGC 6624 (Heise 2010), Grindlay \& Heise (1975)
stumbled over X-ray bursts. Just 10 months before, the first
theoretical paper about thermonuclear burning on neutron stars (NSs)
was published by Hansen \& van Horn (1975). As the observations
showed, the theory needed fine tuning and the first link between
bursts and thermonuclear flashes on NSs was established by Woosley \&
Taam (1976) and Maraschi \& Cavaliere (1977).

From the mid seventies on, X-ray bursts were detected in large
numbers, culminating in a present day count of about 11 thousand,
mostly thanks to recent observations with BeppoSAX WFC (e.g.,
Cornelisse et al.  2003), RXTE (Galloway et al. 2008a), HETE II (e.g.,
Suzuki et al. 2007) and INTEGRAL (e.g., Chenevez et al. 2011).

Actually, the full name of the bursts that we discuss here is Type I
X-ray bursts, to distinguish them from Type II X-ray bursts that are
thought to result from spasmodic accretion in 2-3 sources (e.g., Lewin
et al.  1993), but for brevity we ignore this longer name.

We here briefly review recent developments in X-ray burst research and
discuss expectations for the {\em Monitor of All-sky X-ray Image}
(MAXI). For comprehensive reviews, the reader is referred to Lewin et
al. (1993), Bildsten (1998) and Strohmayer \& Bildsten (2006).

\section{The nature of X-ray bursts}

An X-ray burst is the radiative cooling after a thermonuclear shell
flash occurring just $\sim$1 m beneath the surface of a NS. During
such a shell flash, the accreted mixture of hydrogen and/or helium is
ignited due to a quick pressure build-up at the bottom of the accreted
pile in the very large gravitational field strength on the NS. The
nuclear heating rate increases quicker with temperature than does the
radiative cooling (T$^4$) and a runaway ensues which only dies when
all or most fuel is burnt.  This lasts a fraction of a second, but the
subsequent radiative cooling of the burnt layer lasts longer - of
order 1 min. While the burning layer heats up to $\sim2$ GK (e.g.,
Woosley et al. 2004), the photospheric temperature is limited to a
peak of about $\sim0.03$ GK or $\sim3$~keV, right in the classical
X-ray regime (e.g., Galloway et al. 2008a).

The nuclear chain reaction can become very complex, particularly if
the hydrogen abundance is high at the point of ignition. In that case
rapid proton capture becomes important. It involves hundreds of
isotopes whose decay rates have rarely been measured experimentally in
particle accelerators on earth (e.g., Schatz et al. 2001). Thus, X-ray
bursts are very relevant for nuclear physics (e.g., Davids et
al. 2003; Cyburt et al. 2010).

The range of observational phenomena in X-ray bursts is predominantly
determined by 3 parameters: composition, accretion rate (e.g.,
Fujimoto et al. 1981) and NS spin . This particularly introduces
different X-ray bursting behavior between hydrogen-deficient
ultracompact X-ray binaries (with $P_{\rm orb}<1$~hr) and
hydrogen-rich long-period systems (e.g., in 't Zand et al. 2007) and
large changes during accretion outburst of transients (for a
nice recent example, see Chenevez et al. 2010).

\section{Recent developments}

RXTE and BeppoSAX, launched in 1995/6, provided an enormous enrichment
in the knowledge of the X-ray burst phenomenon.  This pertains to
interesting details, such as long tails in normal X-ray bursts (in 't
Zand et al. 2009), peculiar profiles (e.g., Bhattacharyya \&
Strohmayer 2007; Zhang et al. 2009), and the return of X-ray bursts in
Cir X-1 after 30 years of accretion (Linares et al. 2010), but
foremost in a broader sense as discussed in Sects.  3.1-3.4 below. As
a result, theory improved, on issues such as flame spreading
(Spitkovsky et al. 2002), rotational mixing (Piro \& Bildsten 2008;
Keek et al. 2009), nuclear reaction chains (e.g., Fisker et al. 2008),
convection (e.g., Weinberg et al. 2006), sedimentation (Peng et
al. 2007) and multi-zone 1D simulations of series of bursts (Woosley
et al. 2004).

\subsection{Burst oscillations}

NS spin frequencies were notably lacking for low-mass X-ray binaries
during 30 years of measurements, until Strohmayer et al. (1996)
detected a transient 363 Hz oscillation in 6 X-ray bursts from GX
354-0/4U 1728-34. The strict reproducibility of the (asymptotic value
of the) frequency supported an identification with the NS spin. Many
more sources with such oscillations were quickly found with RXTE, but
it was not until 6 years later (Chakrabarty et al. 2003) that final
proof came for it being due to NS spin, with the simultaneous
detection of a burst oscillation and a millisecond pulsar in SAX
J1808.4-3658. To date, one quarter of all bursters exhibited burst
oscillations, with frequencies between 245 and 620 Hz. The phenomenon
is not completely understood though. Most burst oscillations are
detected in the tails of X-ray bursts, when the NS is presumed to
radiate uniformly. How can the radiation be confined? One explanation
is r-mode oscillations (Heyl 2004; Piro \& Bildsten 2005; Lee \&
Strohmayer 2005; Narayan \& Cooper 2007; Cooper 2008) and another one
is Coriolis force containment (Spitkovsky et al. 2002). A third
obvious idea is that the fuel is magnetically confined. However, the
magnetic dipole field strengths in low-mass X-ray binaries
($\sim10^8$~G) are considered to be insufficient for that.

An interesting discovery was made recently of a peculiar X-ray burster
showing burst oscillations and a pulsar at only 11 Hz (Bordas et
al. 2010; Chenevez et al. 2010; Strohmayer \& Markwardt 2010;
Altamirano et al. 2010) that has very fast series of faint X-ray
bursts, with wait times between bursts down to 5 min (Motta et
al. 2011). The value for $\alpha$ (defined as the fluence in the
accretion flux between bursts divided by the fluence in bursts; see,
e.g., Lewin et al. 1993) is normal, strongly suggesting the X-ray
bursts to be thermonuclear in origin despite the small wait times and
fluences (e.g., Chakraborty et al. 2011; Linares et al. 2011). Fuel
confinement again appears to be a straightforward explanation for this
behavior. Instead of being spread over the entire neutron star, the
accreted matter is confined to a spot (ergo, a pulsar signal) and it
takes less matter, and thus less time, to reach a thick-enough pile
for ignition of a (less energetic) flash. Cavecchi et al. (2011) were
able to exclude r-mode oscillations and Coriolis force containment as
viable explanations for such a slow NS spin, leaving only magnetic
confinement. A magnetic field strength of $B>10^9$~G could be
sufficient for that, and would be consistent with the observed
channeled accretion.  The necessarily larger magnetic field strength
could be related to the small spin frequency, because the Alfv\'{e}n
radius would stretch to slower regions of the accretion disk. Still,
Cavecchi et al. note that this is not a natural explanation for burst
oscillations in many other sources, because there $B$ is unlikely to
be that high. Perhaps the magnetic field is then boosted up during the
burst, an idea put forward by Boutloukos et al. (2010).

\subsection{Long X-ray bursts}

If the ignition is deeper, more mass needs to cool down and the
cooling time is longer. This may range from 10-30 min for so-called
intermediate duration bursts to 1 d for superbursts.  Since the amount
of fuel contained in such thick piles is appropriately larger, the
wait times between bursts is longer - from days to years, depending on
the accretion rate and composition.

{\em Intermediate duration bursts} (in 't Zand et al. 2005; Cumming et
al.  2006, Falanga et al. 2008) are thought to be helium flashes on
cold NSs. The larger pressure of the thicker pile (roughly up to
10~m/10$^{10}$~g~cm$^{-2}$) is thought to compensate for a lower
temperature at ignition. The lower temperature may be the result of
either a low accretion rate, which will decrease crustal heating
through pycnonuclear reactions and electron capture processes, or the
absence of hydrogen in the accreted material, preventing heating by
the CNO cycle, or both. The combination of circumstances seems likely
in many ultracompact X-ray binaries. in 't Zand et al. (2007) propose
from accretion disk theory that persistent low accretion rates can
only occur in ultracompact systems and, thus, identify six new
candidate ultracompact systems from a list of persistent
bursters. However, this diagnostic appears to be not full proof, based
on the detection of hydrogen in the optical spectrum of one such
candidate (Degenaar et al. 2010).

{\em Superbursts}, discovered by Cornelisse et al. (2000), have been
detected 17 times, from 10 sources that also exhibit ordinary X-ray
bursts (e.g., Keek \& in 't Zand 2008b; Kuulkers 2009b). They have
ignition depths of order 10$^2$~m/10$^{12}$~g~cm$^{-2}$. No helium or
hydrogen is thought to survive at those depths. This led Cumming \&
Bildsten (2001) and Strohmayer \& Brown (2002) to suggest carbon as
fuel - a shell-flash analog to core-flash type Ia supernovae. The
carbon is either within the accreted material or produced through
helium/hydrogen burning. In the latter case there must be an intricate
balance with the destruction of carbon through normal X-ray bursts. It
is not clear yet how this balance is reached. The ignition is fairly
close to the presumed NS crust. Thus, superbursts may be good probes
of those crusts (Cumming et al. 2006).  Current theories about crusts
are at odds with understanding superburst recurrence times: they are
measured to be too short as compared with theory (e.g., Keek et
al. 2008a).

\begin{figure*}[t]
\centering
%\psbox[xsize=0.4#1,ysize=0.2#1,rotate=r]
\psbox[xsize=\textwidth]
{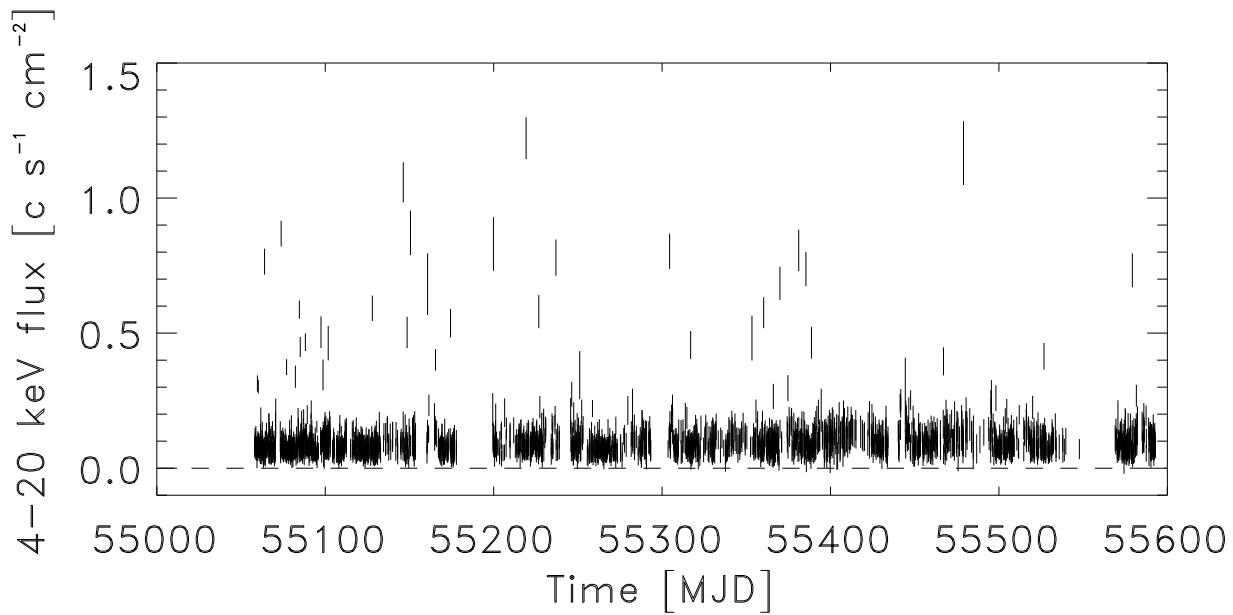}
\caption{MAXI/GSC orbital light curve of GS 1826-24 until Feb 1,
  2011. The fluctuations upwards are indicative of X-ray
  bursts. See Sect.~\ref{maxi1}.\label{fig1826}}
\end{figure*}

\subsection{Short wait times}

At least $y=$10$^8$~g~cm$^{-2}$ of fuel needs to be accreted before
the density threshold for ignition may be reached at the bottom of the
accreted layer. For a NS radius of $10R_{10}$~km, a mass accretion
rate of $10^{-9}\dot{M}_{-9}$ M$_\odot$~yr$^{-1}$ and a pile column
thickness of $10^8y_8$~~g~cm$^{-2}$, it takes $t_{\rm wait}=
5.5R_{10}^2y_8/\dot{M}_{-9}$~hr before that happens\footnote{We ignore
  general relativistic corrections which amount to several tens of
  percents at most.} for accretion onto the complete NS (c.f.,
Sect. 3.1). Therefore, if $t_{\rm wait}$ is less than roughly half an
hour, fuel must have been left from the previous burst. In recent
years this has been observed often.  Boirin et al. (2007) observed the
systematic occurrence of {\em triple} bursts in long ($\sim24$~hr)
uninterrupted observations of EXO 0748-676. The initial burst in these
was always found to have longer tails, indicating that secondary
bursts are always hydrogen-poorer. Keek et al. (2010) confirmed this
result in a larger source sample, after a systematic search for {\em
  multiple} bursts in BeppoSAX and RXTE data based on the MINBAR
database (Galloway et al. 2008b). Keek et al. found multiple bursts in
15 bursters. It is, therefore, a rather common phenomenon. However, no
such bursts were seen from (candidate) ultracompact X-ray binaries,
with a significance of 10$^{-3}$, indicating that hydrogen is a
necessary ingredient. Other features are a shortest wait time of 3.8
min and a {\em quadruple} burst in 4U 1636-536 with the 4 bursts
occurring within a 53 min time span\footnote{4U 1636-53 appears to be
  a rosetta stone of peculiar thermonuclear burning features besides
  multiple bursts: multiple superbursts (Wijnands 2001), mHz
  oscillatory burning (Revnivstev et al. 2001; Heger et al. 2007;
  Altamirano et al. 2008) and odd burst shapes (van Paradijs et al.
  1986; Zhang et al. 2009).}.  Multiple bursts are not understood
yet. It could be related to rotational mixing in these fast rotators.

\subsection{Superexpansion - nova-like shells}

Already in the 1970s X-ray bursts were noted with precursors,
preceding the burst by and lasting for a few seconds (Hoffman et
al. 1978). It was realized (e.g., Hanawa \& Sugimoto 1982) that
super-Eddington luminosities may occur during X-ray bursts, resulting
in photospheric radius expansion. This may be so extreme that the peak
of the black body spectrum moves to lower temperatures, in tandem with
the expansion, and out of the X-ray band introducing the appearance of
a gap in the burst profile (e.g., Tawara et al. 1984). The radial
expansion of the photosphere is by at least a factor of
100. Therefore, this is called {\em superexpansion}, in contrast to
the moderate expansion that is seen in most Eddington-limited
bursts. in 't Zand \& Weinberg (2010) performed a systematic search of
precursors in BeppoSAX, RXTE and published data and found 32 cases
from 8 sources. Interestingly, all sources are (candidate)
ultracompact X-ray binaries. This is a consistent picture:
superexpansion implies the quickest energy release which is typical
for He burning and not for H burning.

The hypothesis about superexpansion is that the luminosity generated
by the nuclear burning is so high that even below the surface of the
NS the Eddington-limit is surpassed and a thick layer is blown away.
For intermediate duration bursts the column thickness may be
$\sim$10$^8$~g~cm$^{-2}$ for an ignition column depth of
$\sim10^{10}$~g~cm$^{-2}$. This shell is optically thick until,
through the expansion, it is diluted so that it becomes optically
thin.  The NS then becomes visible through the shell and the main
burst becomes visible. The rate of expansion can be estimated through
simple black body modeling of the precursor. Velocities have been
measured of up to 10\% of the speed of light. In 3 or 4 cases
(Strohmayer \& Brown 2002; in 't Zand et al. 2011) signatures of
accretion disk disruption by this shell have been detected.

\section{Why are X-ray bursts so fascinating?}

One of the questions in fundamental physics is the behavior of matter
at extreme densities. On the one hand this regime can be probed at
high temperatures with particle accelerators on earth. On the other
hand, for cool temperatures, with NSs (e.g., Paerels et al. 2009).
This is, arguably, the primary driver for X-ray burst research.

NSs harbor the densest bulk matter in the visible universe, with
densities that are higher than that in atomic nuclei, approaching that
of nucleons. The behavior of matter at these densities is described by
quantum chromodynamics (QCD). Currently this theory is incomplete at
high bulk densities due to the uncertain many-body behavior. There is
a dire need of experimental data, ideally the mass and radius of one
NS (e.g., Lattimer \& Prakash 2007; Paerels et al. 2009).

Many X-ray bursts reach luminosities equal to the Eddington limit,
$(2-4)\times10^{38}$~erg~s$^{-1}$. Therefore, they show NS surfaces at
maximum brightness. In principle this makes X-ray bursts ideal tools
to diagnose NSs, particularly to measure their radius which is not
possible in radio pulsars since that radiation arrives from the
magnetosphere instead of the NS itself.

Due to the large compactness of NSs (M/R $\approx$ 0.1~M$_\odot$/km),
just short of that at the Schwarzschild radius
(M/R$\approx$0.3~M$_\odot$/km), space-time is strongly curved around
them. General relativity predicts a gravitational redshift that for
nominal NS parameters (M=1.4~M$_\odot$, R=10~km) amounts to 30\%. If
discrete spectral features were found and identified, this would be a
straightforward measurement. A tentative measurement by Cottam et
al. (2002) spurred many observations with Chandra and XMM-Newton,
including on the same source, but failed to repeat the measurement
(Thompson et al. 2005; Kong et al. 2007; Cottam et al. 2008; Misanovic
et al. 2010).  Nevertheless, this has not been given up yet. in 't
Zand \& Weinberg found evidence of absorption edges in low-resolution
measurements of two superexpansion bursts with RXTE. Such bursts have
not been detected with Chandra or XMM-Newton yet. This is difficult,
because these bursts are so rare.

In the mean time, efforts stepped up to derive NS radii from continuum
burst spectra. In principle the method is straightforward (e.g., van
Paradijs 1979): if the emission is black body radiation, the law of
Stefan-Boltzmann applies: $F = (R_{\rm NS}/d)^2\sigma T^4 $ where
$\sigma$ is the Stefan-Boltzmann constant, $d$ the distance and $F$
the bolometric flux. $F$ and $T$ are measured during bursts, and $d$
can be determined independently, for instance if the source is one of
the 13 or 14 bursters in a globular cluster. $d$ may also be canceled
out if a burst of the same source is seen to reach the Eddington limit
through photospheric radius expansion (see above).  Then $F~d^2=L_{\rm
  Edd}/4\pi$ with $L_{\rm Edd}=4\pi cGM/0.2(1+X)$ and $X$ the hydrogen
abundance. However, as straightforward as this seems, as difficult it
is to infer good constraints on $R_{\rm NS}$. The emission is not
exactly black body due to scattering against hot electrons in the
atmosphere; distances are seldom better determined than 15\% (for
instance, because $X$ is difficult to determine; e.g., Kuulkers et
al. 2003); and the emission may not be isotropic. Effort is underway
to eliminate these systematic uncertainties. The reader is referred to
\"{O}zel et al. (2006), G\"{u}ver et al. (2010), Suleimanov et
al. (2010) and Steiner et al. (2010).

\section{What can MAXI do?}

MAXI (Matsuoka et al. 2009; Sugizaki et al. 2011) scans 95\% of the
2-30 keV X-ray sky each day at a sensitivity of about 15 mCrab. Each
sky position is transited in 40-150 s exposures every 92 min. The
capability is similar to the All-Sky Monitor on RXTE (Levine et
al. 1996), except for a broader energy range (up to 30 instead of 12
keV) and a higher resolution of the photon energy information on the
ground. The sensitivity towards X-ray bursts is, therefore, similar to
RXTE-ASM.

\subsection{An exploratory look at MAXI data}
\label{maxi1}

A good illustration of MAXI's capability on X-ray bursts is provided
by measurements of GS 1826-24. The bursting behavior of this source is
convenient because it exhibits bursts every 3 to 6 hr that are long
with respect to the transit time of the source through MAXI's 3$^{\rm
  o}$ (full-width at zero response) field of view (c.f., Galloway et
al. 2004). Figure~\ref{fig1826} shows the orbital light curve. About
35 spikes can be discerned in this lightcurve. 24 are far from data
gaps and are possibly X-ray bursts.

We have reviewed the orbital data for a few known frequent X-ray
bursters and counted the numbers of possible bursts, see
Table~\ref{tab1}. Many bursts are expected to be shorter than the
transit time. Therefore, the burst signal in 1-orbit accumulations may
be smeared out and the orbital data is not optimum for finding
ordinary X-ray bursts. A typical decay time scale of 10~s implies that
the 1-orbit-averaged signal will be at most 1/4 of the peak flux. The
data needs to be reviewed at a higher time resolution to search for
the typical fast-rise exponential-decay burst profile.  Such a
resolution is not available publicly, because it is non-trivial for
analysis: the data are strongly modulated by the triangular-shaped
transit responses of multiple sources in the FOV. The data also needs
to be reviewed at higher spectral resolution, to search for a cooling
signature that may not be visible in the 3 bands that are at the
moment publicly available.

\begin{table}
\caption[]{Burst counts of MAXI data until Feb 2, 2011.  Exposure time
  (4th column) is simply the number of orbits (3rd column) times 45
  s. The last column provides accumulated times when there are at
  least 2 orbital measurements within 2 hr.\label{tab1}}
\vspace{2mm}
\begin{tabular}{lrrr}
\hline\hline
Object & \# & Orbits & SB expos. \\
       & bursts  &    & (yr)\\
\hline
GS 1826-24 & 24 & 4291 & 0.64 \\
4U 1636-53 & 14 & 4850 & 0.72 \\
Aql X-1$^1$&  7 & 5044 & 0.77 \\
4U 1608-52$^1$&3& 4535 & 0.66 \\
4U 1735-44 &  4 & 3691 & 0.54 \\
4U 1746-37 &  3 & 4280 & 0.64 \\
HETE J1900.1-2455&3&4255&0.63 \\
4U 1724-30 &  2 & 4212 & 0.63 \\
4U 0513-40 &  2 & 3522 & 0.51 \\
SLX 1735-269 & 2& 4477 & 0.65 \\
\hline\hline
\end{tabular}
$^1$Transient with 2 outbursts in MAXI data
\end{table}

\subsection{MAXI on superbursts}

The orbital data are ideally suited to search for superbursts since
the burst duration is much longer than one transit and a burst would
be covered by a number of consecutive transits. The capability is
similar to the RXTE ASM which revealed 8 superbursts in 15 years
(Wijnands 2001; Kuulkers et al.  2002, Remillard \& Morgan 2005;
Kuulkers 2005, 2009b). If one combines all times when two orbital data
points are within 2 hr from each other, for all 10 known
superbursters, the total exposure time is close to 6 yr (see
Table~\ref{tab1}). The average superburst recurrence time is
$2^{+2}_{-0.7}$ yr (in 't Zand et al. 2003), implying that MAXI data
is expected to contain about 3 superbursts. Up to February 2011, no
superburst has been identified yet in MAXI data (Serino,
priv. comm.). The chance probability for detecting none is 5\%.

\section{Future}

There are unexplored niches in X-ray burst research that can be probed
with presently available instrumentation. Chandra and XMM-Newton, with
their spectrographs LETGS, HETGS and RGS, have not yet measured
superexpansion bursts and superbursts, while these are the types of
bursts which have the highest probability for revealing discrete
spectral features, as predicted by theory and suggested by
low-resolution data. RXTE will be operative for at least one more year
and will be able to continue searching for burst oscillations. Just
very recently this brought the surprise of the 11 Hz oscillator. There
is hope that it will break the high-speed record NS spin frequency of
716 Hz (Hessels et al. 2006).  Spin frequencies of 1~kHz or higher are
measurable by RXTE and would rule out some theories for the NS
internal constitution (Lattimer \& Prakash 2007).

The key to significant advancement in future instrumentation is
collecting area, so that burst oscillations can be studied in greater
detail and at lower amplitude, and spectra can be measured more
accurately. A number of proposed future missions with square-meter
collecting area would meet that challenge: IXO, LOFT, AXTAR and
GRAVITAS. With regards to measuring recurrence times of rare X-ray
bursts, it would be worthwhile to have an all-sky monitor with a duty
cycle that is significantly higher than the few percent duty cycles
delivered by for instance MAXI and ASM and with at least moderate
sensitivity, such as proposed on AXTAR (Ray et al. 2010) or MIRAX
(Braga \& Mejia 2006). The already-flying GBM on Fermi has a high duty
cycle and delivers interesting X-ray burst results (Linares et al.,
these proceedings), but has a non-optimum bandpass starting at 8 keV.

%\subsection{Author names and their addresses}

\vspace{1mm}\noindent {\em Acknowledgments.} I am grateful to the SOC
and LOC for inviting me to this very well organized, interesting and
enjoyable workshop. Nobuyuki Kawai and Motoko Serino are thanked for
advice on MAXI data and Anna Watts for useful discussions on burst
oscillations. This research has made use of the MAXI data provided by
RIKEN, JAXA and the MAXI team.

\section*{References}

\re
Altamirano, D., et al. 2008, ApJ, 673, L35

\re
Belian, D. et al. 1972, ApJ, 171, L87

\re
Bhattacharyya, S., \& Strohmayer, T.E. 2007, ApJ, 656, 414

\re
Bildsten, L. 1998, in 'The Many Faces of Neutron Stars', eds.
R. Buccheri, J. van Paradijs, and M. A. Alpar. (Dordrecht: Kluwer
Academic Publishers), p. 419

\re
Boirin, L., et al. 2007, A\&A, 465, 559

\re
Bordas, P., et al. 2010, ATel 2919

\re
Boutloukos, S., et al. 2010, ApJ, 720, L15

\re
Braga, J., \& Mejia, J. 2008, SPIE, 6266, 17

\re
Cavecchi, Y., et al. 2011, ApJL, subm. (arXiv:1102.1548)

\re
Chakrabarty, D. 2003, Nat, 424, 42

\re
Chakraborty, M., et al. 2011, MNRAS, subm. (arXiv:1102.1033)

\re
Chenevez, J., et al. 2010, ATel 2924

\re
Chenevez, J., et al. 2011, MNRAS, 410, 179

\re
Cooper, R.L. 2008, ApJ 684, 525

\re
Cornelisse, R. et al. 2000, A\&A, 357, L21

\re
Cornelisse, R. et al. 2003, A\&A, 405, 1033

\re
Cottam, J. et al. 2002, Nat, 420, 51

\re
Cottam, J. et al. 2008, ApJ, 672, 504

\re
Cumming, A., \& Bildsten, L. 2001, ApJ, 559, L127

\re
Cumming, A., et al. 2006, ApJ, 646, 429

\re
Cyburt, R.H., et al. 2010, ApJS, 189, 240

\re
Davids, B. et al. 2003, PhRevC, 67, id. 065808

\re
Degenaar, N. et al. 2010, MNRAS, 404, 1591

\re
Falanga, M. et al. 2008, A\&A, 679, L93

\re
Fisker, J.L., et al. 2008, ApJS, 174, 261

\re
Fujimoto, M.Y., et al. 1981, ApJ, 247, 267

\re
Galloway, D.K., et al. 2004, ApJ, 601, 466

\re
Galloway, D.K., et al. 2008a, ApJS, 179, 360

\re
Galloway, D.K., et al. 2008b, HEAD abstract 10.21

\re
Grindlay, J., \& Heise, J. 1975, IAUC 2879

%\re
%Grindlay, J., et al. 1980, ApJ, 240, L121

\re
G\"{u}ver, T. et al. 2010, ApJ, 712, 964

\re
Hanawa, T., \& Sugimoto, D. 1982, PASJ, 34, 1

\re
Hansen, C.J., \& van Horn, H.M. 1975, ApJ, 195, 735

\re
Heise, J. 2010, After Dinner Speech, Lorentz Center
workshop 'X-ray bursts and burst oscillations', Leiden,
July 2010 (for slides, see www.lorentzcenter.nl)

\re
Heger, A. et al. 2007, ApJ, 665, 1311

\re
Hessels, J.W.T., et al. 2006, 311, 1901

\re
Heyl, J. 2004, ApJ, 542, L45

\re
Hoffman, J.A., et al. 1978, ApJ, 221, L57

\re
in 't Zand, J.J.M., et al. 2003, A\&A, 411, 487

\re
in 't Zand, J.J.M. et al. 2005, A\&A, 441, 675

\re
in 't Zand, J.J.M. et al. 2007, A\&A, 465, 953

\re
in 't Zand, J.J.M. et al. 2009, A\&A, 497, 469

\re
in 't Zand, J.J.M. et al. 2010, A\&A, 520, A81

\re
in 't Zand, J.J.M. et al. 2011, A\&A, 525, A111

\re
Keek, L. et al. 2008a, A\&A, 479, 177

\re
Keek, L., \& in 't Zand, J.J.M. 2008b, in Proc.
7th INTEGRAL Workshop, Copenhagen, p. 32

\re
Keek, L. et al. 2009, A\&A, 502, 871

\re
Keek, L. et al. 2010, ApJ, 718, 292

\re
Klebesadel, R.W., et al. 1973, ApJ, 182, L85

\re
Kong, A.K.H. et al. 2007, ApJ, 670, L17

\re
Kuulkers, E. 2002, A\&A, 383, L5

\re
Kuulkers, E. et al. 2003, A\&A, 399, 663

\re
Kuulkers, E. 2005, ATel 483

\re
Kuulkers, E. et al. 2009a, A\&A, 503, 889

\re
Kuulkers, E. 2009b, ATel 2140

\re
Lattimer, J.M., \& Prakash, M. 2007, Ph. Rep., 442, 109

\re
Levine, A., et al. 1996, ApJ, 469, L33

\re
Lewin, W.H.G. et al. 1993, SSRv, 62, 223

\re
Linares, M. 2010, ApJ, 719, L84 % Cir X-1

\re
Linares, M. 2011, ApJL, subm. (arXiv:1102.1455) % T5X2

\re
Lee, U., \& Strohmayer, T.E. 2005, MNRAS, 361, 659

\re 
Maraschi, L., \& Cavaliere, A. 1977, in 'X-ray binaries and
compact objects', p. 127

\re
Matsuoka, M., et al. 2009, PASJ, 61, 999

\re
Misanovic, Z., et al. 2010, ApJ, 718, 947

\re
Narayan, R., \& Cooper, R.L. 2007, ApJ, 665, 628

\re
\"{O}zel, F. 2006, Nat, 441, 1115

\re
Paerels, F., et al. 2009, Astro2010 White Paper no. 320 (arXiv:0904.0435)

\re
Peng, F., et al. 2007, ApJ, 654, 1022

\re
Piro, A., \& Bildsten, L. 2005, ApJ, 629, 438

\re
Piro, A., \& Bildsten, L. 2008, ApJ, 663, 1252

\re
Ray, P., et al. 2010, SPIE, 7732, p. 48

\re
Remillard, R., \& Morgan, E. 2005, ATel 482

\re
Revnivstev, M. et al. 2001, A\&A, 372, 138

\re
Schatz, H. et al. 2001, PhRL, 86, 3471

\re
Spitkovsky, A., et al. 2002, ApJ, 556, 1018

\re
Steiner, A.W. et al. 2010, ApJ, 722, 33

\re
Strohmayer, T. et al. 1996, ApJ, ApJ, 469, L9

\re
Strohmayer, T., \& Brown, E. 2002, ApJ, 566, 1045

\re
Strohmayer, T., \& Bildsten, L. 2006, in 'Compact stellar X-ray sources',
ed. W.H.G. Lewin \& M. van der Klis, CUP, p. 113

\re
Strohmayer, T.E., \& Markwardt, C.B. 2010, ATel 2929

\re
Sugizaki, M., et al. 2011, PASJ, in press (arXiv:1102.0891)

\re
Suleimanov, V. et al. 2010, A\&A, subm. (arXiv:1009.6147)

\re
Suzuki, M. et al. 2007, PASJ, 59, 263

\re
Tawara, Y., et al. 1984, PASJ, 36, 845

\re
Thompson, T.W.J. et al. 2005, ApJ, 634, 126

\re
van Paradijs, J. 1979, ApJ, 234, 609

\re
van Paradijs, J. et al. 1986, MNRAS, 221, 617

\re
Weinberg, N.N., et al. 2006, ApJ, 639, 1018

\re
Wijnands, R. 2001, ApJ, 554, L59

\re
Woosley, S.E., \& Taam, R.E. 1976, Nat, 263, 101

\re
Woosley, S.E. et al. 2004, ApJ, 151, 75

\re
Zhang, G., et al. 2009, MNRAS, 398, 368
\label{last}

\end{document}